\documentstyle[11pt,fleqn]{article}

\def\BE{\begin{equation}}
\def\EE{\end{equation}}
\def\BA{\begin{eqnarray}}
\def\EA{\end{eqnarray}}
\def\BAN{\begin{eqnarray*}}
\def\EAN{\end{eqnarray*}}
\def\BN{\begin{enumerate}}
\def\EN{\end{enumerate}}
\def\BI{\begin{itemize}}
\def\EI{\end{itemize}}

\def\DS{\displaystyle}
\def\LAN{\langle}
\def\RAN{\rangle}

\author{V.N.Gorbachev, A.I.Zhiliba}
\title{Transfer formalism for quantum optics problems }
\begin{document}
\maketitle
\begin{abstract}
Consistent quantum formalism based on the localized basis of 
the Wannirer functions in Heisenberg and Schrodinger pictures 
to describe propagation of electromagnetic field 
in a three dimensional media including diffraction is presented.
In the  Schrodinger picture the Fokker-Planck equation for the 
Glauber-Sudarshan  quasiprobability and corresponding Langevin 
equations are given. As result the space-time description is 
obtained by a simple changing variables in the temporal master 
equation of the field. Using this formalism it is shown that the existence of 
integrals of motion in the propagation of light in a medium 
under the condition of nondegenerated parametric and two-photon 
interactions results in amplification of modes when 
nonclassical properties of the light are conserved.
Quantum propagation of light in a linear medium 
taking into account the diffraction is considered and its solution is found. 

\end{abstract}
\newpage
\section{Introduction}
When considering statistics of the light the usual approaches are based on 
the temporal evolution of the electromagnetic field. It is natural for the problems  
when the spatial behavior can be neglected as for the high-Q optical cavity scheme 
inside which one the temporal feature is important. By contrast the spatial behavior 
plays the key role in propagation of the light through media or distributed system 
that can be not considered as a point one. In these systems new states of the light 
named the spatially squeezed states arise \cite{1}.
The quantum formalism for distributed systems has been developed in a number 
of works. One-dimensional approaches were presented in \cite{2}.
Theory for continuous variables in the Heisenberg picture was given in 
\cite{3}, where the continual coherent states and squeezed states of the field have been 
introduced and theory of the light detection in free space has been presented.
With the use of these approaches evolution of the light statistics has been examined for 
the three-photon \cite{4} and four-photon \cite{5} parametric interaction in transparent media. 
The multi-photon processes was discussed in \cite{6}.

In this paper we present a quantum formalism which enables to consider 
propagating of the light in a three-dimensional medium with 
diffraction to be taken into account.
It is based on a localized basis of the Wannirer functions, and allows the 
conversion of the temporal description to space-time description by a simple change of variables. 
Indeed it is the way used in solid state physics for transition from collective to local variables \cite{7}. 
The main feature of the local description in the Wannirer basis is that the problem turns out 
to be multi-particle when the local field oscillators interact  among themselves  even in  
free space. From the physical point of view this is a transfer of excitation or the light 
propagation process. In other hand the interaction of local oscillators produces 
the coupled equations which as we show  have the form of the transfer equations.
Our approach is formulated in both the Heisenberg and Schrodinger pictures where 
the Fokker-Planck equations for the field quasiprobability are derived 
together with the corresponding Langevin equations. 

To describe interaction of light with atoms the adiabatic illumination of fast atomic variables is often used to obtained a closed equation of field or master equation \cite {8}, \cite {9} that is the start point to analyze statistics of the light. The simple recipe of how to get the local description allows to obtained the transfer equation immediately from the temporal master equation 
missing all steeps of the derivation procedure.

The potential of the presented formalism is illustrated by two problems. 
In the first one the statistics of the light is considered for the propagation 
in a one-dimensional medium with two-photon and parametric interaction. 
Here within the framework of local description an interesting peculiarity 
due to the integrals of motion arises. The existence of the integrals makes 
it possible to establish the main features of the statistics transformation 
immediately without going into the solution of the dynamic problem \cite{10}.
It has been shown by this way that enhancement of light and the conservation 
of nonclassical properties can be possible. In the second problem the propagation 
of the light in a linear three-dimensional medium with diffraction is considered 
and the solutions have been found.

\section{Wannirer basis. Local operators}
In one-dimensional normalization space $L$ plane waves
$\varphi _{k}(x) = (1/\sqrt{L})\exp (ikx)$, where
$\Delta k=2\pi/L$,
are forming a complet orthonormal basis.
It is used for a standard representation of the electromagnetic field 
strength  operator, where the operators of photon creation and annihilation 
$a^{\dagger}_{k}, a_{k}$ are arising with commutational relations 
\BE
[a_{k}, a^{\dagger}_{k'}]=\delta _{kk'}
\EE
Operators $a^{\dagger}_{k}, a_{k}$ describe the creation and annihilation 
of photons of a wave vector $k$ in the whole space $L$.
These operators may be named collective, because they are responsible 
for the excitation of the whole volume. For the local description of the 
electromagnetic field instead of plane waves we use Wannirer functions 
known in solid state physics \cite{7}, which are packets of plane waves
\BE
w_{m}(x-l)=\frac{1}{\sqrt{N}} \sum _{k \sim m} \exp (-ikl)
\varphi _{k}(x)
\label{21}
\EE
the wave vectors $k \sim m$ are lying in a band
$
m-\pi/a \leq k < m+\pi/a
$. 
Here each band $m$, or zone, corresponds to the partition  of one-dimensional 
space $L$ into $N=L/a$ cells, which positions are determined as $l=na$. 
In k-space the centers of so defined bands are separated by an interval  
$\Delta m=2\pi /a$.
The terms in (\ref{21}) are forming a geometrical 
progression, so summation is possible. Then it easily verified that for 
large $N$ the Wannirer functions are localized in a cell with a coordinate $l$ 
in region  $\sim a$. The Wannirer functions defined in accordance with (\ref{21}) 
form the complet orthonormal set 
\BA
\sum _{ml} w_{m}(x-l) w_{m}^{\ast}(x'-l)&=&\delta (x-x')\\
(w_{m}(x-l), w_{m'}^{\ast}(x-l'))&=&\delta_{mm'}\delta_{ll'}
\EA
where the scalar product is defined by the integration 
over the whole space $L$.

Wannirer basis serves as a framework for the introduction of local operators 
of creation and annihilation
$a^{\dagger}_{ml}, a_{m}(l)$:
\BE
\sum _{k} \varphi _{k}(x)a_k =\sum _{ml}a_{m}(l)w_{m}(x-l)
\label{23}
\EE
From (\ref{23}) it follows that the operators are coupled by a unitary transformation
\BE
a_{m}(l) = \sum _{k}C^{\ast}_{mk}(l)a_{k}
\label{25}
\EE
\BE
a_{k} = \sum_{ml}C_{mk}(l)a_{m}(l)
\label{26}
\EE
where
\BE
C_{mk}(l)=\frac{1}{\sqrt{N}}\sum _{k'\sim m} \exp (-ik'l)\delta_{k'k}
\label{261}
\EE
For local operators the following commutational relations are valid:
\BE
[a_{m}(l), a^{\dagger}_{m'}(l')]=\delta _{mm'}\delta _{ll'}
\label{28}
\EE
The relations \label{28} (\ref{23}) allow to interpret them 
as operators of creation and annihilation of a photon in a point $l$ 
in the vicinity $\sim a$. As a result local field oscillators described 
by the introduced operators are defined in the space $L$. From here on we 
will refer the packets of plane waves forming the local operators as local 
modes with wave number $m$ and with a width $\Delta \nu_{m}=c/a$. Strictly 
speaking these packets are not modes, because there are no physical reasons 
for their distinctions, however the term seems to be convenient.

The equation (\ref{23}) is a starting point in the formalism of quantum 
transfer theory, which may be formulated in different pictures. 
From the operational point of view the unitary transformations (\ref{25}) and 
(\ref{26}) following from (\ref{23}) mean that the transition from the nonlocal 
to local description and vice versa is accomplished by the change of variables.

The specific feature of local description is its many-particle character, 
where the local oscillators are interacting already in the free space. 
This interaction describes the excitation transfer or light propagation 
process, which determine the structure of equations of motion, where 
chains of the BBGKI type (Bogolyubov, Born, Green, Kirkwood, Ivon) are 
arising resulting finally in  the propagation equations. 

Consider the case of free space, for which the field evolution is defined 
only by a Hamiltonian $H_{0}=\sum _{k} \hbar \omega_{k}a^{\dagger}_{k}a_{k}$.
$\omega_{k}=ck$.  Let the operator $a_{k}$ in (\ref{23}) be defined in the 
Heisenberg representation, i.e. satisfying the equation 
$\partial a_{k}/\partial t=i\hbar^{-1}[H_{0}, a_{k}]$. 
To find the equation for the local Heisenberg operator 
\BE
a_{m}(l,t)=\frac{1}{\sqrt{N}}\sum _{k \sim m}a_{k}(t)\exp (ikl)
\label{33}
\EE
let us make the differentiation with respect to time
Assuming the size of the space cell $a$ is small, $a\ll L$, and $a \to 0$, $l$ 
may be considered as a continuous space coordinate.  
In this approximation 
\BE
k\exp(ikl)=-i\frac{\partial}{\partial l} \exp (ikl)
\label{35}
\EE
as result it follows that the transfer equation for the local operator is
\BE
\left(
\frac{\partial}{\partial t}+c\frac{\partial}{\partial l}
\right)
a_{m}(l,t)=0
\label{36}
\EE
This equation is valid in a coarse space scale with a characteristic 
size $a$ providing the fulfillment of (\ref{35}), in this case the 
replacement $\delta_{ll'}\to a\delta(l-l')$ is possible.

For a many-particle problem (\ref{36}) may be represented as set of coupled  
equations due to the interaction of local operators. The Hamiltonian $H_{0}$ 
appears to be nondiagonal with respect to the indices $ll'$. The equation 
of motion takes the form
\BE
\frac{\partial}{\partial t}a_{m}(l)=i\hbar^{-1}[H_{0}; a_{m}(l)]
=-i\sum_{l^{`}}\Omega _{m}(l,l^{`})a_{m}(l`)
\label{361}
\EE 
the coupling constant 
\BE
\Omega_{m}(l,l^{`})=\frac{c}{N}\sum_{k\sim m}k\exp(ik(l-l^{`}))
\label{362}
\EE
links the oscillator in the point $l$ with all neighbors. 
However, due to its $\delta$ -shape, the coupling appears to be significant 
only for two adjacent oscillators. In other words, in the free space 
the interaction between local oscillators is described in a coarse scale 
by a derivative with respect to $l$:
\BE
\frac{\partial}{\partial l}a_{m}(l)
=\frac{i}{c}\sum_{l^`}\Omega_{m}(l,l^`)
a_{m}(l^`)
\label{39}
\EE
the equation (\ref{39}) is obtained 
by differentiation of (\ref{33})  with respect to $l$ with taking into account (\ref{35})
Consider the commutator of local operators taking for different times.
For the free evolution, when the dynamics  of operators is determined 
by the Hamiltonian $H_{0}$	
\BE
[a_{m}(l,t); a^{\dagger}_{m^`}(l,t+\tau)]=
\delta_{mm`}\exp (icm\tau)\frac{1}{\Delta\nu_{m}}\delta_{a}(\tau)
\label{310}
\EE
the function (2.17) 
\BE
\delta_{a}(\tau)=
\frac{\Delta\nu_{m}}{N}\frac{\sin (\pi\tau\Delta\nu_{m})}
{\sin (\pi\tau\Delta\nu_{m}N^{-1})}
\label{D}
\EE
$N \gg 1$ has a sharp maximum for $\tau \to 0$. Its weight is concentrated in 
the vicinity of the order of  $a/c=\Delta\nu_{m}^{-1}$, so it may be 
considered as a delta-function when $\Delta t \geq a/c$. 
Notice that $\delta_{a}(0)=\Delta\nu_{m}$. The occurence of the scale temporal 
delta-function is connected to the use of coarse space scale with a characteristic 
size $a$, where the time interval will be $\Delta t \geq a/c$.	

Let us introduce the interaction picture. Consider the slowly changing 
part or envelope in the local operator $a_{m}(l)$: 
\BE
a_{m}(l,t)=A_{m}(l,t)\exp (-i\omega _{m}t+iml)
\label{320}
\EE 
where $\omega _{m}=cm$, Then the unitary transformations in (\ref{25}) and (\ref{26}) 
will take the form
\BE
A_{m}(l,t)=\frac{1}{\sqrt{N}}\sum_{k\sim m}a_{k}(t)
\exp \{-i(\omega_{k}-\omega_{m})t+i(k-m)l\}
\label{211}
\EE
\BE
a_{k}(t)=\frac{1}{\sqrt{N}}\sum_{l}A_{m}(l,t)
\exp \{i(\omega_{k}-\omega_{m})t-i(k-m)l\}
\label{212}
\EE
the evolution of the operators $a_{k}$ and $A_{m}(l,t)$ is defined 
only by the Hamiltonian of interaction $V$
\BA
\frac{\partial}{\partial t}a_{k}(t)&=&i\hbar ^{-1}[V(t),a_{k}(t)]
\\
\left(
\frac{\partial}{\partial t}+c\frac{\partial}{\partial l}
\right)
A_{m}(l,t)&=&
i\hbar ^{-1}[V(t),A_{m}(l,t)]
\label{213}
\EA

\section{Three - dimentional case. Quasioptical approximation}

Consider a normalized volume $L^{3}$ with a number of cells
$N=N_{1}N_{2}N_{3}$,$N_{1}=N_{2}=N_{3}=l/a$, where for 
simplicity the cell was chosen to be of cubic shape. Then in the initial
formulas  the evident replacements will be:
$
k \to  {\bf k}$, $m \to {\bf m}$,$l \to {\bf l}$, 
$x\to {\bf r}$.
For the local operator in the  interaction picture
the following expression should be written instead of (\ref{211}):
\BE
{\bf A_{m}(l},t)=\frac{1}{\sqrt{N}}
\sum_{\bf k \sim m} {\bf a_{k}}(t)
\exp (-i(\omega_{\bf k}-\omega_{\bf m})t+
{\bf (k-m)l})
\label{101}
\EE
where the dispersion law has the form 
$\omega_{\bf q}=c\sqrt{(\bf q,q)}$, ${\bf q=k.m}$. 
Let us find the equation of motion 
for local operators. Similarly to the one-dimensional case let us
differentiate (\ref{101}) with respect to time. 
Assuming ${\bf l}$ to be a continuously changing variable, we will find
instead of (\ref{35})
\BE
\sqrt{(\bf q,q)}\exp (i{\bf ql})
\approx \left(-i\frac{\partial}{\partial l_{z}}-
\frac{1}{2q_{z}}\left( \frac{\partial^2}{\partial l_{x}^2}
+\frac{\partial^2}{\partial l_{y}^2}\right) \right) \exp (i{\bf ql})
\label{102}
\EE
When summing over ${\bf k \sim m}$ we replace 
$k_{z} \to m \approx m_{z}$
This approximation corresponds to a quasiplane wave
of frequency  $\omega_{m}=cm$, propagating 
along the $z$-axis.
In as much as a coarse spatial scale was introduced, for which 
vector ${\bf l}$ may be considered as a continuous function of the 
coordinates, let us  make a replacement  
$
{\bf l}\to {\bf r}(x,y,z)
$.
As a result the equation of motion takes the form
\BE
\left(\frac{\partial}{\partial t}
+c\frac{\partial}{\partial z}
-i\frac{c^{2}}{2 \omega_{m}}
\left(
\frac{\partial^{2}}{\partial x}
+\frac{\partial^{2}}{\partial y}
\right)
\right) {\bf A_{m}(r)}=
-i\hbar^{-1}[V, {\bf A_{m}(r)}]
\label{1020}
\EE
where $V$ is the Hamiltonian of intraction. The equation following 
from (\ref{1020}) for the free field ($V=0$) is well known in 
the classical theory. It describes the light propagation with 
diffraction taken into account in a quasioptical approximation.

In the equation (\ref{1020}) the Hamiltonian $V$ describing the 
electromagnetic field interactions with the medium should be expressed 
in terms of local operators. As an example consider the Hamiltonian of 
the light interaction with atoms in the dipole approximation, which 
is often a basis for a variety of problems. It has the form
\BE
V(t)=-i\sum_{A,{\bf k}}
\sqrt{\frac{\hbar\omega_{{\bf k}}}
{2\varepsilon_{0}L^{3}}}{\bf a_{k}}\exp(-i\omega_{{\bf k}}t+i{\bf kr}_{A})
d_{A}(t)
+h.c.
\label{321}
\EE
The equation (\ref{321}) is given in the interaction picture, 
where $d_{A}$ is the operator of the dipole moment for an atom located 
at a point ${\bf r}_{A}$. Changing to local operators with the aid of 
(\ref{212}), where the three-dimensionality should be taken into account, 
we will use the following approximations.
Let the packet or the local mode interacts with the atom as a whole. 
It means that inside the band $\Delta\nu_{m}$ all frequencies 
$\omega_{k}\approx \omega_{m}$. Replace the atom
position ${\bf r}_{A}$ by the position of the cell where this atom is located. 
Then the summation over the atoms may be divided into a sum over the 
cells ${\bf l}$ containing atoms and a sum over the atoms inside the cell. 
Suppose $d_{l}$ is the operator of the atomic dipole moment in the cell ${\bf l}$, 
then the equation (\ref{321}) will take the form 
\footnote[1] 
{details of these and following calculations are given in \cite{11}}
\BE
V(t)=-i\sum_{{\bf ml}}\sqrt{\frac{\hbar\omega_{{\bf m}}}
{2\varepsilon_{0}a^{3}}}{\bf A_{m}(l})\exp(-i\omega_{{\bf m}}t+i{\bf ml})
d_{l}+h.c.
\label{322}
\EE
Here a new normalization volume $a^3$ appears, while the summation is
performed only over the cells containing atoms.
The obtained Hamiltonian describes the elementary interactions of 
the local field oscillators or photons in cells $l$ with atoms located inside. 
As a result chaiging to 
local oscillators in the interaction Hamiltonian (\ref{321}) reduces to an
ordinary replacement: 
${\bf a_{k}} \to {\bf A_{m}(l})$, ${\bf k} \to {\bf m}$, 
${\bf r} \to {\bf l}$.

In a number of cases effective interaction operators obtained, for example, 
by unitary transformations of the starting Hamiltonian ((\ref{321}), are used for 
the description of multi - photon processes. 
Given below are two effective interaction operators for two - photon and parametric 
interactions in the presentation of local operators:
\BA
H_{2}&=&\sum_{{\bf l}}\sum_{1,2}
f_{12}
{\bf A_{m_{1}}(l)A_{m_{2}}(l)}S_{l}
\exp(i(\omega_{21}-\omega_{{\bf m}_{1}}-\omega_{{\bf m}_{2}})t)
+h.c.
\label{325}
\\
H_3&=&\sum_{{\bf l}}\sum_{1,2,3}
G_{123}({\bf l})
{\bf A_{m_{1}}(l)A_{m_{2}}(l)A^{\dagger}_{m_{3}}(l)}
\exp(-i(\omega_{{\bf m}_{1}}+\omega_{{\bf m}_{2}}
-\omega_{{\bf m}_{3}})t)
\label{326}
\\
\nonumber
&+&h.c.
\EA
where $f_{12}$, 
$G_{123}=g_{123}\exp (i({\bf m_{1}+m_{2}-m_{3})l})$ - 
are the coupling constants. The Hamiltonian $H_{2}$ describes 
two-photon interaction of the modes having frequencies 
in the region of two-photon resonance:
$\omega_{21} \approx \omega_{{\bf k_{1}}}+\omega_{{\bf k_{2}}}$. 
The operator $S_{l}=|2\RAN _{l}\LAN 1|$ corresponds to the transition 
of atoms located at the point $l$ from the lower to the upper working level. 
The process of parametric interaction of three waves in a transparent 
medium conforms to the Hamiltonian $H_{3}$.

\section{Integral of motion and statistics of the light}
The integrals of motion may appear in the problems of light propagation 
in a medium, for which it is natural to use local description. The existence 
of the integrals enables to examine some peculiarities of the light 
statistics transformation without going into the solution of dynamic equations. 
Consider as an example two nonlinear processes: two-photon and parametric 
interactions, which are described by the effective Hamiltonians (\ref{325}) 
and (\ref{326}). Both processes are multimode, involving all pairs and 
triplets of modes having frequencies related as
\BA
\label{331}
\omega_{m_{1}}+\omega_{m_{2}}&=&\omega_{21}
\\
\label{332}
\omega_{m_{1}} + \omega_{m_{2}}&=&\omega_{m_{3}}
\EA
Here after we restrict ourselves by the case of one-dimensional medium 
assuming the light propagating along the $z$-axis.

The pairs of modes $m_{1}$ and $m_{2}$ connected by the conditions (\ref{331}) 
and (\ref{332}) and (4.31) will be referred to as conjugated. For them 
the difference of the photon number operators 
\BE
I=n_{1}(l)-n_{2}(l)
\label{115}
\EE
where $n_{j}=A^{\dagger }_{m_{j}}(l)A_{m_{j}}(l)$, $j=1,2$ 
commutate with the Hamiltonian
\BE
[H_{2,3}, I]=0
\label{116}
\EE
Consequently $I$ and any function of the
form $f(I)$ are integrals of motion. For a problem with boundary 
conditions $l$ will be considered as a continuous coordinate replacing  
$
l \to z
$.
Introducing the new variables and assuming the light propagating along the $z$-axis:
$
t'=t-z/c, z'=z
$.
Then by virtue of (\ref{116}) the difference of the number of photons in conjugated 
modes is conserved for any point of the medium:
\BE
I(z,t)=I(0,t-z/c)
\label{117}
\EE
With the use of the obtained operator integrals of motion it is possible 
to find the peculiarities of the transformation of the mutual correlation 
of conjugated modes for the propagation in a medium. Thus all statistical 
properties described by the correlation functions of differented intensity, 
e.g. of the form $\LAN (I(z,t))^{p}I((z,t+\tau))^{q}\RAN = K^{(p,q)}(z,\tau)$ 
are conserved in virtue of (\ref{117}). From the standpoint of observation of 
direct interest is the lowest order correlation function and its Fourier image
\BE
K(z,\Omega )=\int^{\infty}_{-\infty}
\LAN I(z,t)I(z,\tau)\RAN \exp (i\Omega
\tau)d\tau
\label{001}
\EE  
\begin{figure}
\unitlength=1.00mm
\special{em:linewidth 0.4pt}
\linethickness{0.4pt}
\begin{picture}(110.37,42.34)
\put(49.33,16.67){\line(1,0){21.33}}
\put(70.66,12.34){\framebox(10.00,7.67)[cc]{$D_{2}$}}
\put(81.00,16.67){\line(1,0){12.33}}
\put(50.33,37.67){\line(1,0){21.33}}
\put(71.66,33.34){\framebox(10.00,7.67)[cc]{$D_{1}$}}
\put(82.00,37.67){\line(1,0){12.33}}
\put(94.00,26.67){\circle{10.02}}
\put(94.00,37.67){\line(0,-1){6.00}}
\put(94.00,16.67){\line(0,1){5.00}}
\put(99.33,26.34){\line(1,0){8.00}}
\put(94.00,26.34){\makebox(0,0)[cc]{--}}
\put(107.00,33.00){\makebox(0,0)[cc]{$i_{-}$}}
\put(55.33,42.34){\makebox(0,0)[cc]{1}}
\put(55.33,20.00){\makebox(0,0)[cc]{2}}
\put(66.00,38.00){\line(-4,1){6.33}}
\put(65.67,37.67){\line(-3,-1){6.67}}
\put(60.00,18.67){\line(3,-1){5.67}}
\put(59.67,14.33){\line(5,2){5.33}}
\put(109.00,26.33){\circle*{2.75}}
\end{picture}
\caption{}
\label{r2}
\end{figure}

It is possible to measure the function (\ref{001}) in a scheme with 
two photodetectors (Fig.\ref{r2}), where the fluctuation spectrum of difference 
photocurrent $i^{2}(\Omega)$ is observed. For this scheme
\BE
i^{2}(\Omega)=\eta \LAN n_{1}+n_{2}\RAN +\eta^{2}K_{N}(\Omega) 
\label{002}
\EE
 
Here the first independent of frequency term is the shot noise. 
Index $N$ indicates normal ordering of the field operators
to describe a detector that responds at the absorption of photon. The rate of registration 
$\eta=q\Delta\nu_{m}$, 
where $q$ is the quantum efficiency of the detectors,
which for simplicity are 
assumed to be equal. The presence of the local mode $\Delta\nu_{m}=c/a$ 
in the expression for the photocurrent spectrum follows from the formulas 
of detection given in the presentation of local operators. This may be 
elucidated as follows. A wide-band photodetector is 
needed to describ the light detection. 
Therefore to registrate the local mode in front of the 
detector we have to place an optical filter of a bandwidth $\Delta\nu_{m}$ . 
Then $\Delta\nu_{m}$ that determines 
the rate at which the photons fall on the detector
is the bandwidth of the scheme.

Assuming that at detector the local operators commute as 
free field operators (\ref{310})  
let us to write  the correlation function (\ref{001}) 
in the normal and time-ordered form 
$K(\Omega )=K_{N}(\Omega )+\LAN n_{1}+n_{2}\RAN /\Delta\nu_{m}$. 
Then taking into account the motion integral one finds
\BE
K_{N}(z,\Omega)+\frac{1}{\Delta \nu_m}\LAN n_1+n_2 \RAN=
K_{N}(0,\Omega)+\frac{1}{\Delta \nu_m}\LAN n_{10}+n_{20} \RAN
\label{118}
\EE
where $\LAN n_{10}\RAN $ and $\LAN n_{20} \RAN $   
are the input photon number.  
As result the expression for the photocurrent or noise spectrum (\ref{002}) 
takes the form 
\BE
i^{2}(z,\Omega)=\Delta \nu_{m}\LAN n_1+n_2 \RAN q(1-q)
+\Delta \nu_{m}\LAN n_{10}+n_{20} \RAN q(q-1)+
i^{2}(0,\Omega)
\label{1120}
\EE
where
\BE
i^{2}(0,\Omega)=\Delta \nu_{m}q\LAN n_{10}+n_{20} \RAN+
(\Delta \nu_{m} q)^{2}K_{N}(0,\Omega)
\EE
is the input noise.

It is seen from (\ref{1120}) that for an ideal photodetector ($q=1$) 
the input and output noise spectra of the light 
are equal.
This means that the correlation of conjugated modes is conserved 
during propagation in the medium. 
Suppose that the input state of the modes 
is nonclassical, so it is a quantum correlation for which 
the shot noise is suppressed in bandwidth $\Delta\nu_{m}$
by a factor of $1-q$:
\BE
i^{2}(0,\Delta \Omega)=\Delta \nu_{m}q\LAN n_{10}+n_{20} \RAN
(1-q)
\label{003}
\EE
The state of the light resulting in (\ref{003}) produces, for 
example, by optical parametric amplifier (OPA). Then the level of the 
shot noise suppression in the output is unchanged:
\BE
i^{2}(z,\Delta \Omega)=\Delta \nu_{m}q\LAN n_{1}+n_{2} \RAN
(1-q)
\EE
This means that the quantum property of the light in the medium is conserved. 
Moreover in the medium with two-photon or parametric interaction the 
conjugated modes may be amplified, at that the observed initial 
correlation, in particular the quantum correlation, is conserved. 
This specific feature determines the properties of spontaneous radiation. 
Thus, if input is the vacuum state, 
then $i^{2}(z,\Delta \Omega)=0$ at the output. 
This means that the conjugated modes arising in 
spontaneous radiation have nonclassical correlation, that results in 
a suppression of shot noise. Indeed  all these properties are 
following only from the existence of integrals of 
motion in systems expanded in space.

\section{Local quasiprobabilities }
In previous items the Heisenberg picture was introduced for 
local operators. In the Schrodinger picture the evolution is determined by 
the density matrix of electromagnetic field $\rho$, which may be connected 
to c-number functions $P(\{\alpha_{k}\},s)$ called s-ordered quasiprobabilities. 
They arise in the density matrix expansion over operators 
$\Delta(\{\alpha_{k}\},s)$, which are Fourier-images of s-ordered 
displacement operators forming the complet set:
\BE
\Delta(\{\alpha_{k}\}, s)=\frac{1}{\pi}\int \{d^{2}\beta_{k}\}
\prod _{k}
\exp \{s|\beta_{k}|^{2}+
(\beta^{\ast}_{k}(\alpha_{k}-a_{k})-h.c.)\}
\EE

$\{d^{2}\beta_{k}\}=\prod_{k}d^{2}\beta_{k}$.
Then the $s$-ordered quasiprobability is determined by the expression 
\BE
P(\{\alpha_{k}\}; s)
=Sp\left( \Delta(\{\alpha_{k}\}, s)\rho\right)
\label{PD}
\EE

Using the replacements $a_{k}\to a_{m}(l)$ in the above formulas we obtain 
the local quasiprobability $P(\{\alpha_{m}(l)\}; s)$. In the following we 
will restrict ourselves by the case s=1 corresponding to the normal ordering 
of the field operators which is described by the Glauber-Sudarshan 
quasiprobability $P(s=1)=P$. This function arises in the density matrix 
expansion over coherent states or diagonal representation. From the given expressions it 
is possible to obtain the relation between local and nonlocal quasiprobabilities. 
Similary the case of operators, it is possible to proceed from one distribution function 
to another by the change of variables:
\BA
P(\{\alpha_{m}(l)\})&\longleftrightarrow&  P(\{\alpha_{k}\})\\
\{\alpha_{m}(l)\}&\longleftrightarrow& \{\alpha_{k}\}
\label{45}
\EA
where the variables are related by the expression of the (\ref{25}) and (\ref{26}).

In the Schrodinger representation the peculiarities of the local 
description specified by the many-particle character of the problem 
become apparent to the same extent as in the Heisenberg representation. 
Here the transfer equations occur, which are formally 
equivalent to the BBGKI chains for the partial distribution functions. 
Of all the hierarchy of the distribution functions in the present case 
the following two types of one-particle quasiprobabilities are of interest. 
They arise from the function $P(\{\alpha_{m}(l)\})$, 
of all modes of all local oscillators. Averaging over all oscillators except 
for the chosen one we will find the distribution function 
$P(\alpha_{\{m\}}(l))$ for all modes of one local oscillator located at 
a point $l$. The averaging of $P(\alpha_{\{m\}}(l))$ over all 
modes except for one results in a function $P(\alpha_{m}(l))$ describing one 
mode of one local oscillator. 

For a free evolution the introduced one-particle distribution 
functions $P_{1}=P(\alpha_{\{m\}}(l))$, $P(\alpha_{m}(l))$ obey the 
following transfer equation:
\BE
\left(
\frac{\partial}{\partial t}+c\frac{\partial}{\partial l}\right) P_{1}
=0
\label{69}
\EE
In (\ref{69}) the derivative over $l$ describes the interaction of local 
oscillators or excitation transfer. 
Similarly to (\ref{361}) it may be presented in the form 
\BE
c\frac{\partial}{\partial l}P(\alpha_{m}(l))
=-h_{m}(l,l)P(\alpha_{m}(l))-\sum_{l'\neq l}\int d^{2}\alpha_{m}(l')
h_{m}(l,l')P(\alpha_{m}(l),\alpha_{m}(l')
\label{699}
\EE
where 
$
h_{m}(l,l')=i\Omega_{m}(l,l^`)
(\partial/ \partial \alpha_{m}(l))
\alpha_{m}(l^`)+c.c.
$
is the differential operator of free evolution with 
Hamiltonian 
$H_{0} \leftrightarrow \sum_{mll'} h_{m}(l,l')$.

\section{Fokker - Planck and Langevin equations}
An approach based on the master equation for the quasiprobability 
$P(\{\alpha_{k}\})$ in the Fokker-Planck approximation is often used 
to describe the statistical properties of light. In the problems 
of the light interaction with atoms the master equation for the 
electromagnetic field can be obtained by an adiabatic 
elimination of atomic variables. It is not necessary to derive the field 
equation once again in term of the local description, because 
one may immediately use the change of variables (\ref{45}). 

As an example consider the interaction of light with 
a two-level system in the lowest approximation for which the 
processes of the linear amplification or absorption type occur. 
Such a medium is described by a linear susceptibility,   
$
\kappa(\omega)=|d|^{2}(\omega_{0}-\omega -i\gamma)^{-1}
$
where $d$  and $\omega_{0}$
is the dipole moment and the frequency of the atomic transition,
$\gamma$  is the decay rate or transverse relaxation
The form of the field equation obtained by the adiabatic 
elumination of the fast atomic variables is well known (see, \cite{9}):
\BE
\frac{\partial}{\partial t}P(\{\alpha_{k}\})=
\sum_{k}\left(A(k)\frac{\partial}{\DS \partial \alpha_{k}}\alpha_{k}+
Q(k)\frac{\partial^{2}}{\DS \partial \alpha_{k}\partial
\alpha_{k}^{\ast}}+c.c.\right)P(\{\alpha_{k}\})
\label{71}
\EE
Here the coefficients are defined by the linear susceptibility 
and populations of the upper $N_{2}$ and lower $N_{1}$ levels:
\BAN
A(k)&=&\epsilon_{k}(N_{1}-N_{2})Im ~\kappa(\omega_{k})
-i\epsilon_{k}(N_{1}+N_{2})Re ~\kappa(\omega_{k})\\
Q(k)&=&\epsilon_{k}N_{2}Im ~\kappa(\omega_{k})\\
\epsilon_{k}&=&\hbar^{-2}(\hbar \omega_{k}/2\varepsilon_{0}L^{3})
\EAN
When changing the variables we will assume that the 
wave packet forming the local mode interacts as a single whole. Then inside 
the band $k\sim m$ it is possible to neglect the dispersion of all harmonics. 
This means that  
$
A(k)\approx A(m)
$,
$
Q(k)\approx Q(m)
$.
This approximation enables us to write at once the 
transfer equation for one-particle quasiprobability, say 
$P(\alpha_{\{m\}}(l))=P_{1}$ in the form
\BE
\left(
\frac{\partial}{\partial t}+c\frac{\partial}{\partial l}
\right)
P_{1}=
\sum_{m}\left( A(m)
\frac{\partial}{\DS \partial \alpha_{m}(l)}\alpha_{m}(l)+
Q(m)\frac{\partial^{2}}{\DS \partial \alpha_{m}(l)\partial
\alpha_{m}^{\ast}(l)}\right)P_{1}
+ c.c.
\label{74}
\EE
Notice that the structure of differential operators in right- 
hand site is 
the same in (\ref{74}) and (\ref{71}), and the whole of the 
passage procedure to the local description reduces to an addition 
of a derivative over $l$ to the left-hand side of the equation. 

The equation of quasiprobability (\ref{71}) is in agreement 
whith the following Langevin equations:
\BA
\frac{\partial}{\partial t}\alpha_{k}&=&
-A(k)\alpha_{k}+f_{k}
\label{81}\\
\frac{\partial}{\partial t}\alpha_{k}^{\ast}&=&
-A^{\ast}(k)\alpha_{k}^{\ast}+f_{k}^{\dagger}
\nonumber
\EA
where the correlator of random forces is defined by a diffusion coefficient
\BE
\langle f_{k`}(t)f_{k}^{\dagger}(t+\tau)\rangle =
2Q(k)\delta_{kk`}\delta(\tau)
\label{82}
\EE
In as much as the Langevin variables here are associated with 
the diagonal representation they correspond to the normally ordered 
averages of the field operators  
$
\LAN a^{\dagger}_{k}(t)a_{k}(t+\tau) \RAN=
\LAN \alpha_{k}^{\ast}(t)\alpha_{k}(t+\tau)\RAN
$.

In the local description the main moment in the formulation of 
Langevin equations is the finding of the random force correlator. 
To this end we use  changing of variables of the form (\ref{211}), (\ref{212}), 
where $A_{m}(l)\to\alpha_{m}(l)$ $a_{k}\to\alpha_{k}$, thereby 
we introduce the local Langevin variable $\alpha_{m}(l)$. Then the 
approximation $A(k)\approx A(m)$, $Q(k)\approx Q(m)$ used in writing 
down the equation for local quasiprobability leads to Langevin equations
\BA
\left(
\frac{\partial}{\partial t}+c\frac{\partial}{\partial l}
\right)
\alpha_{m}(l,t)&=&
-A(m)\alpha_{m}(l,t)+f_{m}(l)
\label{85}\\
\left(
\frac{\partial}{\partial t}+c\frac{\partial}{\partial l}
\right)
\alpha_{m}^{\ast}(l,t)&=&
-A^{\ast}(m)\alpha_{m}^{\ast}(l,t)+f_{m}^{\dagger}(l)
\nonumber
\EA
For the used variables there follows an expression for a 
random source
\BE
f_{m}(l,t)=\frac{1}{\sqrt{N}}\sum_{k\sim m}f_{k}
\exp (-i(\omega_{k}-\omega_{m})t+i(k-m)l)
\EE
From this we will find that the correlator of random 
sources as well as in (\ref{82})  is defined by the  diffusion coefficient
\BE
\langle f_{m`}(l`,t)f_{m}^{\dagger}(l,t+\tau)\rangle =
2Q(m)
\delta_{mm`}\delta_{ll`}\delta(\tau)
\label{86}
\EE 
As to (\ref{86}) it should be noted that during the transition to a 
local description with a coarse space scale a scaling time $\delta$-function 
should be used defined as in (\ref{D}) and the $l$ magnitude should 
be considered as a continuously coordinate. This implies that in 
(\ref{86}) the replacements take place 
$
\delta(\tau) \to \delta_{a}(\tau)$,
$
\delta_{ll`} \to a\delta (l-l`)
$.
As a result the correlator of random sources in local Langevin equations has the form
\BE
\langle f_{m`l`}(t)f_{ml}^{\dagger}(t+\tau)\rangle =
2Q(m)
\delta_{mm`}a\delta(l-l`)\delta_{a}(\tau)
\label{87}
\EE
In comparison with the ordinary description in (\ref{87}) there arises 
a space delta-function due to the local correlation of oscillators in space.

\section{Diffraction in linear medium}
The differential operator in (\ref{1020}) becomes non-Hermitian due 
to diffraction effects, so the corresponding equation for quasiprobability 
becomes very complicated. However it is possible to use the Langevin formulation. 
In this way for the case of a linear medium considered above the change of 
variables in nonlocal Langevin equations (\ref{85}) leads to Langevin equations 
which in the three-dimensional case have the form
\BA
\left(\frac{\partial}{\partial t}
+c\frac{\partial}{\partial z}
-i\frac{c^{2}}{2 \omega_{m}}
\left(
\frac{\partial^{2}}{\partial x}
+\frac{\partial^{2}}{\partial y}
\right)
\right)\alpha_{m}&=&
-A(m)\alpha_{m}+f_{m}({\bf r},t)
\label{103}\\
\left(\frac{\partial}{\partial t}
+c\frac{\partial}{\partial z}
+i\frac{c^{2}}{2 \omega_{m}}
\left(
\frac{\partial^{2}}{\partial x}
+\frac{\partial^{2}}{\partial y}
\right)
\right) \alpha^{\ast}_{m}&=&
-A^{\ast}(m)\alpha_{m}^{\ast}+f_{m}^{\dagger}({\bf r},t)
\nonumber
\EA
where the random force correlator is defined as
\BE
\langle f_{m`}({\bf r'},t)f_{m}^{\dagger}({\bf r},t+\tau)\rangle =
2Q(m)
\delta_{mm`}a^{3}\delta({\bf r-r`})\delta_{a}(\tau)
\label{104}
\EE
Here $Q(m)$ is the diffusion coefficient, which may be obtained
from the corresponding  (\ref{103}) but nonlocal  Fokker-Plank equation
(\ref{71}): $Q(m)\approx Q(k)$.

The equations (\ref{103}) are linear in field amplitudes, so they are 
easily integrated. Let us denote the transverse radius-vector 
$
{\bf s} =(x,y)
$
and introduce a travelling coordinate system 
$
t'=t-z/c$, $d z'=z$.
Then for the given 
conditions at the boundary one finds
\BA
\alpha_{m}(z,{\bf s},t)&=&
\exp (-\frac{A}{c}z) \int d^{2}{\bf s}_{1}\ \alpha_{m}
(0,{\bf s}_{1},t-z/c)
\ U(0{\bf s}_{1}|z{\bf s})
+W(z,{\bf s},t-z/c)\\
W&=&\frac{1}{c}\int_{0}^{z}dz_{1}\exp(-\frac{A}{c}(z-z_1))
\int d^{2}{\bf s }_{1} f_{m}(z_{1},{\bf s}_{1},t-z/c)
U(z_{1}{\bf s}_{1}|z{\bf s})
\EA
where the Green function is defined as
\BE
U(z_{1}{\bf s}_{1}|z{\bf s})=
-i\frac{m}{2\pi (z-z_{1})}\exp \left(im \frac{({\bf s}-{\bf
s}_{1})^{2}}{z-z_{1}}\right)
\EE
Here $ \alpha_{m}(0,{\bf s}_{1},t-z/c)$ is the transverse field distribution 
in the input, $A=A(m)$

Consider the properties of a random source $W$, defining its correlation function
\BE
\LAN W(z,0,t) W^{\dagger}(z,{\bf s},t+\tau)\RAN=
\frac{2}{c^{2}}
Qa^{3}\delta_{a}(\tau)\int^{z}_{0}\exp \left(-\frac{A+A^{\ast}}{c}
(z-z_{1})\right)D(z-z_{1})dz_{1}
\EE
where the value
\BE
D=\left\lbrack \frac{m}{2\pi (z-z_{1})}\right\rbrack ^{2}
\exp \left(-i\frac{ms^{2}}{2(z-z_{1})}\right)
\int d^{2}{\bf s}_{1}\ \exp \left( i\frac{m}{z-z_{1}}{\bf s_{1}s}\right)
\EE
describes the diffraction effects. Since the diffusion coefficient 
$Q$ is determined by the number of atoms at the upper level $N_2$ the 
transverse sizes of the medium where the light propagates should 
be taken into account. In this way the limits of integration over 
${\bf s}_{1}$ in $D$ are defined. Thus 
$
D=\delta ({\bf s})
$
if the medium is not limited in the transverse direction.
This means that the noise source is $\delta$-correlated and the noise 
is white. Taking into consideration the finiteness of the size in transverse 
direction changes the situation. Thus in a medium confined by a cylinder 
with a radius $R$
\BE
D= \frac{mR}{2\pi(z-z_{1})s}
J_{1}\left(\frac{msR}{z-z_{1}}\right)
\exp \left(-i\frac{ms^{2}}{2(z-z_{1})}\right)
\EE
where $J_{1}$ is the Bessel function.


\begin{thebibliography}{99}
\bibitem{1}
I.V. Sokolov, M.I. Kolobov. Zh.Eksp.Teor.Fiz. 96, 1945, (1989)\\
M.I. Kolobov, I.V. Sokolov. Phys. Lett., 140, 101, (1989)
\bibitem{2}
Yu.M.Golubev Zh.Eksp.Teor.Fiz.65, 466, (1973)\\
I. Abram. Rhys. Rev. A 35, 4661, (1987)\\
C.M. Caves, D.D. Crouch. J.Opt. Soc. Am. B 4, 1553, (1987)\\
T.A. Kennedy, E.M. Wright. Phys. Rev. A 38, 212, (1988)
\bibitem{3}
K.J. Blow, R. Loudon, S.J.D. Phoenix. Phys. Rev. A 42, 4102 (1990)
\bibitem{4}
Yu.M.Golubev, V.N. Gorbachev. Zh.Eksp.Teor.Fiz, 95, 475, (1989)\\
V.N. Gorbachev, A.I. Trubilko. Zh.Eksp.Teor.Fiz, 103, 1931,(1993)\\
V.N. Gorbachev, A.I. Trubilko. Opt. and Spektrosk., 80, 301,(1996)
\bibitem{5}
K.J. Blow, R. Loudon, S.J.D. Phoenix. J. Opt. Soc. Am. B 8, 1750,
(1991)\\
K.J. Blow, R. Loudon, S.J.D. Phoenix. Phys. Rev. A 42, 8064 (1992)
\bibitem{6}
V.N. Gorbachev, A.I. Trubilko. Zh.Eksp.Teor.Fiz, 102, 1441, (1992)\\
J.R.Jeffers, N. Imoto, R. Loudon. Phys. Rev. A 47, 3346, (1993)
\bibitem{7}
C. Kittel. Quantum Theory of Solids. New-York - London, 1963\\
A.S.Davidov. The solid state theory.Moskva.Nauka. 1976\\
\bibitem{8}
M.Scully, W.E. Lamb. Phys.Rev. 159, 208, (1967)\\
H.Haken. Laser Theory. Encyclopedia of Physics 2nd edn vol.
XXY/2c (Berlin: Springer). 1984.\\
L.A. Lugiato, F. Casagrande, I. Pizzuto. Phys.Rev. A 26, 3438, (1982).\\
\bibitem{9}
V.N. Gorbachev, A.I. Zhiliba. Quant. Opt. 5, 193, (1993).
\bibitem{10}
V.N. Gorbchev, A.I. Trubilko. Optics and Spektroscopy, 84, 
879, (1998)
\bibitem{11} V.N. Gorbachev, A.I. Zhiliba. Transfer equation in statistical
nonlinear optics. Reprint-98, Tver state University. 1998.
\end{thebibliography}
\end{document}